\documentclass[a4paper, oneside, twocolumn, notitlepage, 10pt]{extarticle_ecoc}
\usepackage{ecoc}
\usepackage{tikz}
\usepackage{pgfplots}
\usepackage{pgfplotstable}
\usepackage[nolist]{acronym}
\usepackage{orcidlink}
\usepackage[justification=justified,singlelinecheck=false]{caption}
\begin{document}
\selectlanguage{english}    


\title{Receiver-Informed Transmitter for 200Gbps C-Band Direct-Detection PON With FFE-Free ONU}


\author{
    Safiya Dabwan\textsuperscript{(1)}, Geraldo Gomes\textsuperscript{(1)},
    Stylianos Sygletos\textsuperscript{(1)}
}

\maketitle                  


\begin{strip}
    \begin{author_descr}

        \textsuperscript{(1)}  Aston Institute of Photonic Technologies, Aston University, Birmingham, UK, B4 7ET,
        \begin{center}\textcolor{blue}{\uline{sdabw20@aston.ac.uk}} \end{center}

    \end{author_descr}
\end{strip}

\renewcommand\footnotemark{}
\renewcommand\footnoterule{}


\begin{strip}
    \begin{ecoc_abstract}
We introduce and numerically demonstrate receiver-informed transmitter (RITX) waveform design for 200~Gb/s PAM-4 C-band direct-detection PON. Using 50~GHz optoelectronics, RITX achieves N1/E1 loss classes with FFE-Free ONU, enabling 40~km reach and 48.1\%/66.7\% network-wide complexity savings at 1:64/1:128 splits.

    \end{ecoc_abstract}
\end{strip}

\section{Introduction}
Future very-high-speed PONs (VHSPs) are expected to deliver 200Gbps per wavelength, yet the cost and complexity constraints of massive-scale ONU deployment strongly favour direct detection (DD) as the preferred reception scheme. In parallel, O-band congestion driven by successive PON generations, combined with interoperability requirements, motivates the exploration of alternative bands, such as the C-band.
C-band DD operation in VHSPs is fundamentally limited by square-law photodetection, which removes optical phase, introduces signal-to-signal beating interference (SSBI), and combined with chromatic dispersion (CD) causes frequency-selective power fading \cite{zhu2020a}. While increasingly sophisticated receiver-side processing has been explored \cite{Yi2019}, a more practical approach shifts CD compensation to the transmitter using complex-valued FIR filters pre-equalisation and an IQ-MZM, forming a hybrid IQ-DD architecture \cite{rizzelliHighspeedPONSolutions2026a}. This DSP centralisation strategy aligns naturally with PON's asymmetric cost structure, as OLT complexity is shared across all ONUs in the network\cite{Zhang2017}.

Prior work exploiting DSP centralisation at the OLT within an IQ-DD architecture has proven effective for 100~Gbps Class-N1 transmission over 20~km using PAM-4 and 25G-class O/E, requiring a 20-tap adaptive FFE and a square-root (SQRT) nonlinear mitigation function at the ONU \cite{torresferrera100GbpsCDPON2022}. Extending to 200~Gbps using PAM-8 and 50G-class O/E has also been demonstrated for Class-N1 at 20~km \cite{wang200GbpsPONDownstream2022}, albeit requiring a SQRT function and a highly complex Volterra nonlinear equaliser (VNLE) at the ONU. Both approaches, however, rely on fixed analytical waveform shaping, leaving substantial residual ISI and nonlinear distortions to be handled at the ONU.

End-to-end (E2E) learning has emerged as a powerful optimisation framework for optical communications \cite{OsheaHoydis,karanovEndtoEndDeepLearning2018a,nielsenEndtoEndLearningTransmitter2025}, yet existing work has largely focused on point-to-point links and joint transmitter--receiver optimisation. This assumption is incompatible with PON's asymmetric architecture, where receiver complexity is replicated across every ONU in the network. For DD-PON, the central challenge is whether receiver feedback can instead optimise the \emph{transmitter alone} to synthesise waveforms matched to the post-detection channel.

In this paper, we propose a receiver-informed transmitter (RITX) framework for 200~Gb/s PAM-4 C-band DD-PON, in which OLT-side FIR filters are optimised via ONU feedback to synthesise waveforms matched to the post-detection channel at the ONU, enabling an FFE-free ONU. RITX achieves 200~Gb/s over 20~km for Class-N1 and Class-E1, extends network reach to 40~km for Class-N1, and reduces network-wide equalisation complexity by 48.1\% and 66.7\% at split ratios of 1:64 and 1:128, respectively, all with with an FFE-free ONU. 

\begin{figure}[t!]
    \centering
    \includegraphics[width=1\columnwidth]{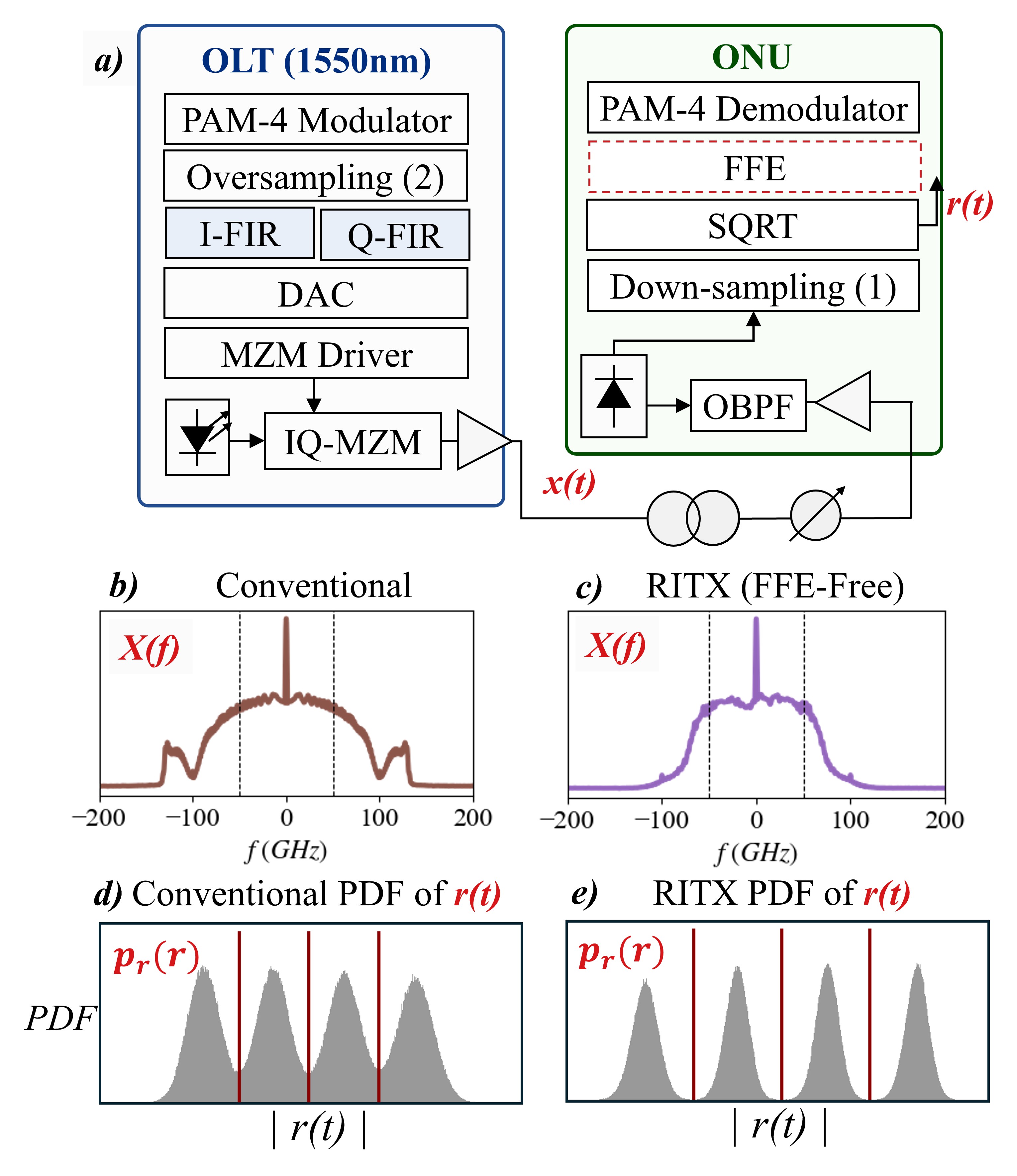}
    \caption{(a) Block diagram of the simulated 200~Gb/s PAM-4 C-band DD-PON system. In the conventional scheme, ONU reception includes SQRT and FFE, whereas in the proposed RITX scheme the transmitter waveform is optimized to enable FFE-free ONU operation. (b,c) Conventional and RITX transmitted spectra, with the apparent spectral chopping near the band edge for the first case. The RITX achieves improved confinement inside the usable 50\,GHz band. (d,e) Probability density function (PDF) of the detected signal amplitude $r(t)$ for the conventional and RITX schemes, respectively. The conventional waveform leaves stronger residual ISI, leading to greater overlap between PAM-4 levels, whereas RITX reduces residual ISI and improves level separation.}        
    \label{fig:fig1}
\end{figure}

\section{Principle and Simulation Setup}

The proposed receiver-informed transmitter (RITX) framework centralises waveform design at the OLT by adapting complex-valued transmitter filters using feedback from the ONUs. Because ONUs can experience different fibre lengths, the effective end-to-end direct-detection channel is not identical for all users. However, ONUs with similar propagation distances exhibit similar post-detection channel responses and can therefore be served by the same filter pair. Based on this observation, the served distance range is partitioned into spatial clusters, each assigned one pair of complex-valued FIR filters at the OLT.

To distinguish between filter adaptation and service, RITX operates in two modes. In \textit{training mode}, the OLT transmits pilot symbols and receives analogue feedback from a single corresponding instructor ONU representing each spatial cluster. This feedback, together with the known pilot, is used to optimise the cluster-specific transmitter filters by minimising the mean-squared error (MSE) over a digital twin of the link. Thus, RITX does not simply invert chromatic dispersion; it learns a launch waveform that is favourable for the detected ONU signal under the composite DD-PON channel. In \textit{service mode}, the OLT selects the filter pair associated with the destination ONU cluster and applies it before optical modulation, thereby pre-compensating the dominant impairments at the shared transmitter and enabling simplified ONU reception without receiver-side FFE.

To evaluate the concept, we numerically simulate a downstream 200~Gb/s PAM-4 C-band link using the IQ-DD architecture, shown in Fig.~\ref{fig:fig1}(a). At the OLT, PAM-4 symbols are oversampled to 2 samples/symbol and processed by a pair of complex-valued FIR filters. In the conventional reference system \cite{wang200GbpsPONDownstream2022}, the filter coefficients are fixed and derived from the analytical CD model of~\cite{Chagnon2019}, assuming prior knowledge of the accumulated $D\!\cdot\!L$ product, and combined with a pre-emphasis filter to compensate transmitter bandwidth roll-off. In the proposed RITX system, the same analytical solution is used only for initialisation, after which the filter coefficients are adapted using receiver feedback at the training mode. The key distinction of RITX is that it moves beyond analytical CD pre-compensation, to performing holistic waveform design informed by the receiver feedback. The RITX-trained filters therefore synthesise a launch waveform that is matched to the effective post-detection DD-PON channel, rather than only inverting the analytical CD response. As illustrated in Fig.~\ref{fig:fig1} (b)-(c), this leads not only to improved confinement of the transmitted waveform within the usable 50~GHz bandwidth, but also to flatter passband. This results into better separation of the detected PAM-4 amplitude distributions, indicating reduced residual ISI as demonstrated by the probability density functions of the received signal before FFE Fig.~\ref{fig:fig1} (d)-(e).

RITX training is performed over a differentiable TensorFlow~2.6 digital twin, enabling backpropagation from the ONU-side error to the OLT-side filter coefficients. Using 4096 pilot symbols, the filters are optimised by Adam (learning rate $10^{-2}$) over 100 epochs. Their real and imaginary outputs then drive the in-phase and quadrature branches of the IQ-MZM for full-field waveform shaping. Where digital twin modelling is impractical, gradient-free alternatives such as CMA-ES used \cite{masaadExperimentalDemonstration4Port2024} or measurement-based surrogate models \cite{hernandezExperimentalDemonstrationEndtoEnd2025} can be employed.

The IQ-MZM sinusoidal transfer characteristic is included to account for transmitter nonlinearity, with the in-phase and quadrature arms biased at quadrature and null, respectively. The electrical bandwidth limitations of the modulator and receiver are modelled by second-order Gaussian low-pass filters with 3~dB bandwidth of 50~GHz. Semiconductor optical amplifiers (SOAs) are placed at the OLT and ONU as booster and pre-amplifier stages, respectively, with gains of 7~dB and 11~dB and a noise figure of 7.5~dB. Fibre propagation is simulated by solving the nonlinear Schr\"odinger equation with the symmetric split-step Fourier method, including chromatic dispersion, Kerr nonlinearity, and attenuation. The carrier wavelength is 1550~nm, with $D=17$~ps/(nm$\cdot$km), $\gamma=1.3$~(W$\cdot$km)$^{-1}$, and $\alpha=0.2$~dB/km. ODN loss conditions are emulated with optical attenuation, with N1-class operation used as the main reference case unless otherwise stated.

At the ONU, a fourth-order Gaussian optical bandpass filter suppresses out-of-band amplified spontaneous emission noise before square-law photodetection. The photodetector is modelled with responsivity 0.7~A/W and input-referred noise density 15~pA/$\sqrt{\mathrm{Hz}}$, while shot and thermal noise are represented as additive Gaussian processes. Receiver DSP operates at 1 sample/symbol and includes square-root function for SSBI mitigation \cite{wang100GbpsCBand2021a}. An adaptive FFE is included only where required by the evaluated scheme: in the conventional system, its tap count is swept to determine the minimum value needed to satisfy the target performance, whereas in the RITX case the FFE-free condition is directly assessed. Performance is measured by direct error counting over $2\times10^6$ bits using a pre-FEC BER threshold of $10^{-2}$~\cite{zhangProgressITUTHigher2020b}.


\section{Results and Discussion}
\begin{figure*}[t]
    \centering
    \includegraphics[width=1\textwidth]{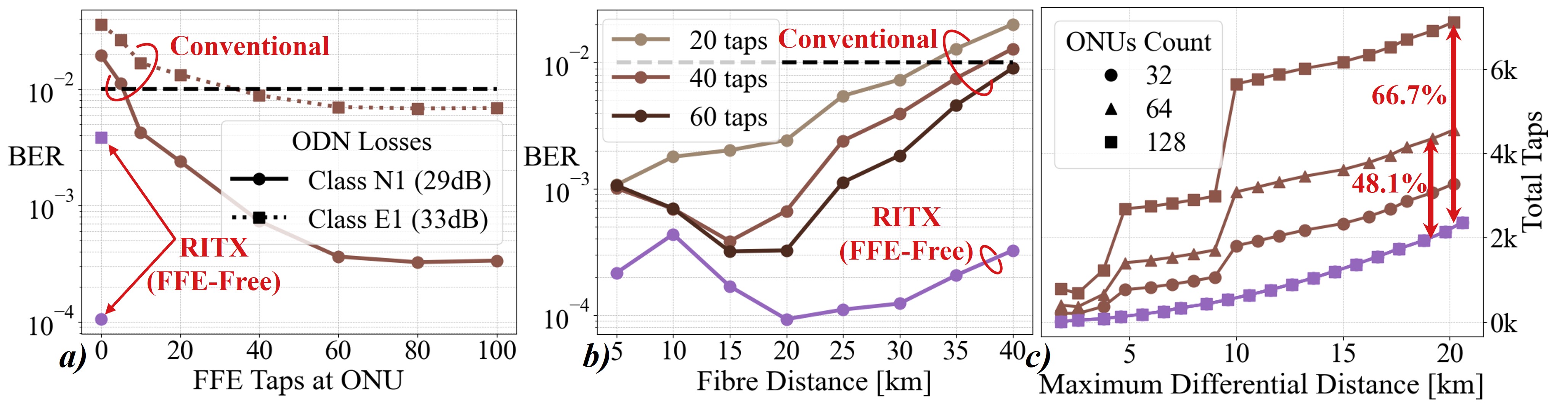}
     \caption{(a) Minimum ONU-side FFE tap count required by the conventional scheme to meet the pre-FEC BER threshold, compared with the proposed RITX system operating with FFE-free, under N1- and E1-class ODN conditions. (b) BER versus fibre distance under N1-class for RITX and for the conventional scheme with different FFE tap counts. (c) Total network-wide equalization tap count versus maximum differential distance for split ratios of 1:32, 1:64, and 1:128.}
    \label{fig:fig2}
\end{figure*}

In Fig.~\ref{fig:fig2}(a) we compare the ONU-side FFE requirement of the conventional scheme with the proposed RITX system. RITX satisfies the pre-FEC BER threshold of $10^{-2}$ with \emph{zero} ONU FFE taps under both N1- and E1-class ODN conditions. In contrast, the conventional reference requires substantial receiver equalisation to approach the same operating point. Under E1-class losses, at least 40 FFE taps are required to cross the FEC BER threshold. While under N1-class losses, approaching the performance margin achieved by RITX requires the conventional system to have more than 60 taps. Thus, RITX removes the need for ONU-side linear equalisation, whereas fixed analytical pre-compensation does not.

Figure~\ref{fig:fig2}(b) evaluates the achievable reach under N1-class ODN conditions. The proposed RITX system maintains BER below $10^{-2}$ over the complete span of 40~km, corresponding to the extended differential-distance requirement defined for future PON \cite{ITU-T2021}. By comparison, the conventional scheme requires a large number of FFE taps to approach this reach and still operates with little or no BER margin at the longest distances. This confirms that RITX not only eliminates ONU equalisation, but also improves the robustness of 200~Gb/s DD-PON transmission against accumulated propagation impairment.

Both the conventional and RITX systems require a bank of cluster-specific transmitter filters at the OLT to accommodate the variation in accumulated CD across ONUs at different differential distances. Following the approach of \cite{torres-ferrera100GbpsPON2020}, in which filter banks collectively cover the full 20~km span by each serving a specific sub-range, we partition the differential-distance range into spatial clusters, each served by one pair of OLT FIR filters and represented during training by its corresponding instructor ONU. The cluster width is determined by the distance span over which a given filter pair can still meet the $10^{-2}$ BER target. Because RITX more tightly tailors each waveform to the full post-detection channel, its achievable span per filter pair is generally narrower than that of the conventional analytical scheme, resulting in 20 OLT filter pairs for RITX compared to 18 for the conventional system. Despite this marginal increase in transmitter-side filter count, the complete elimination of receiver FFE yields a substantially lower total tap count at the network level.

The deployment trade-off is quantified and illustrated in Fig.\ref{fig:fig2}(c), by computing the total equalisation tap count for three PON split ratios (1:32, 1:64, 1:128) over a maximum differential distance $DD_\mathrm{max}$ of 0.6--20.2\,km. The total tap count is defined as $T_\mathrm{total} = T_\mathrm{TX} + N_\mathrm{ONU} T_\mathrm{RX} $,
where $T_\mathrm{TX}$ is the aggregate number of OLT-side transmitter filter taps across all spatial clusters required to cover the assumed differential-distance $DD_\mathrm{max}$, $N_\mathrm{ONU}$ is the number of ONUs in the network, and $T_\mathrm{RX}$ is the minimum ONU-side FFE tap count required per user to serve the widest-case distance within each cluster. For RITX, $T_\mathrm{RX}=0$ by design.

For a 1:64 split-ratio network, RITX achieves a 48.1\% reduction in total tap count relative to the conventional system (4,560 taps reduced to 2,368 taps). For a future 1:128 network, this grows to a 66.7\% reduction (7,120 taps versus 2,368 taps). The scaling advantage of RITX arises because receiver-side taps accumulate linearly with ONU count in the conventional system, while the RITX transmitter tap count remains fixed and independent of the ONU-count.
\section{Conclusion}
We have demonstrated that receiver-informed transmitter (RITX) waveform design enables 200 Gb/s PAM-4 C-band DD-PON with a fully FFE-free ONU, 40 km N1-class reach, and 48.1\%/66.7\% reduced network-wide equalisation complexity at 1:64/1:128 splits than analytical pre-compensation. By centralising adaptive waveform synthesis at the OLT, RITX provides a scalable route to low-cost VHSP ONUs, with experimental validation and upstream operation as natural next steps.

\section{Acknowledgements}
This work was partially supported by the Royal Academy of Engineering under the Research Chairs and Senior Research Fellowships scheme.

\printbibliography
\end{document}